\documentclass[10pt,conference]{IEEEtran}
\usepackage{graphicx,cite,amssymb,amsmath,xcolor,subfig}
\IEEEoverridecommandlockouts      % This command is only needed if you want to use the \thanks command
\newtheorem{defi}{Definition}
\newtheorem{thm}{Theorem}

\newtheorem{corr}{Corollary}
\newtheorem{fact}{Fact}

\newcommand{\be}{\begin{equation}}
\newcommand{\ee}{\end{equation}}
\newcommand{\ben}{\begin{equation*}}
\newcommand{\een}{\end{equation*}}
\newcommand{\mc}{\mathcal}

\newcommand{\mcb}{\mathcal{B}_{M,L}}
\newcommand{\bfs}{\mathbf{S}}
\newcommand{\bfy}{\mathbf{Y}}
\newcommand{\bfx}{\mathbf{X}}

\newcommand{\bfxh}{\mathbf{\hat{X}}}

\newcommand{\abs}[1]{\lvert#1\rvert}
\newcommand{\norm}[1]{\lVert#1\rVert}

\begin{document}
% paper title
\title{\vspace{10pt} Sparse Regression Codes for  Multi-terminal\\ Source and Channel Coding}
\author{
\authorblockN{Ramji Venkataramanan}
\authorblockA{Dept. of Electrical Engineering\\
Yale University, USA \\
Email: ramji.venkataramanan@yale.edu}
\and
%\authorblockN{Antony Joseph}
%\authorblockA{Dept. of Statistics\\
%Yale University, USA\\
%Email: antony.joseph@yale.edu}
%\and
\authorblockN{Sekhar Tatikonda}
\authorblockA{Dept. of Electrical Engineering\\
Yale University, USA \\
Email: sekhar.tatikonda@yale.edu}
}
\maketitle
\vspace{-20pt}
\begin{abstract}
  We study a new class of codes for Gaussian multi-terminal source and channel coding. These codes are designed using the statistical framework of high-dimensional linear regression and are called Sparse Superposition or Sparse Regression codes. Codewords are linear combinations of subsets of columns of a design matrix. These codes were introduced by Barron and Joseph and shown to achieve the channel capacity of AWGN channels with computationally feasible decoding. They have also recently been shown to achieve the optimal rate-distortion function for Gaussian sources. In this paper, we demonstrate how to implement random binning and superposition coding using sparse regression codes. In particular, with minimum-distance encoding/decoding it is shown that sparse regression codes attain the optimal information-theoretic limits for a variety of multi-terminal source and channel coding problems. \end{abstract}

\section{Introduction} \label{sec:intro}
 Among the important outstanding problems in network information theory is developing codes for various multi-terminal source and channel models that are provably rate-optimal with computationally efficient encoding and decoding algortihms. The introduction of deep  ideas such as superposition \cite{CoverBC72}, random binning \cite{SW73} and auxiliary random variables \cite{Wyner75,WynerZiv,GelfandPinsker} has led to a sharp characterization of information-theoretic limits for several network problems. However, until recently, even the best feasible codes for these problems fell short of these limits.

There have been some recent breakthroughs that begin to bridge this gap. Polar codes
were the first codes with computationally feasible encoding algorithms that were shown to provably attain the
information-theoretic limit for discrete-alphabet symmetric sources and channels \cite{Polarcc, Polarrd, Polarwzgp, AbbeTelatar}. Spatially coupled ensembles have recently been shown to achieve the capacity of binary-input symmetric-output channels with belief propagation decoding \cite{spatialCoup12}. There
are many important communication settings where the source or channel alphabet is inherently
continuous, notably Gaussian sources and AWGN channels. Elegant techniques such as lattice
coding have been proposed for continuous-alphabet source and channel coding \cite{Zamir02, EyForney93, ErezZ04}, but these
rate-optimal coding schemes do not have feasible encoding and decoding algorithms.
%The divergence between information-theoretic results and code-construction has been even more pronounced in network communication problems.

Recently a class of codes called Sparse Superposition Codes or Sparse Regression Codes (SPARC) was introduced by Barron and Joseph \cite{AntonyML,AntonyOMP,AntonyFast} for communication over the AWGN channel. In \cite{AntonyFast}, it was shown that SPARCs achieve the AWGN channel capacity with a computationally feasible decoding algorithm. SPARCs have also been shown to attain the optimal rate-distortion function of Gaussian sources with feasible algorithms \cite{KontSPARC, RVGaussianRD12, RVGaussianFeasible}. In this paper, we show that the sparse regression framework can be used to design  feasible codes for various Gaussian multi-terminal source and channel models.

\begin{figure}[t]
\centering
\subfloat[\small{Compressing $X$ with decoder side-information $Y$}]{
\includegraphics[width=2.5in]{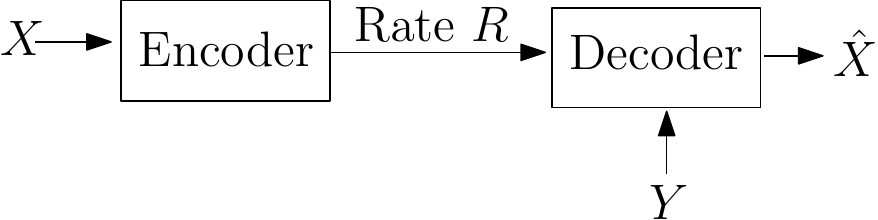}
\label{fig:wz}
}
\vspace{10pt}
\subfloat[\small{Communicating over channel $P(Y|X S)$ with state $S$ known at the encoder}]{
\includegraphics[width=2.9in]{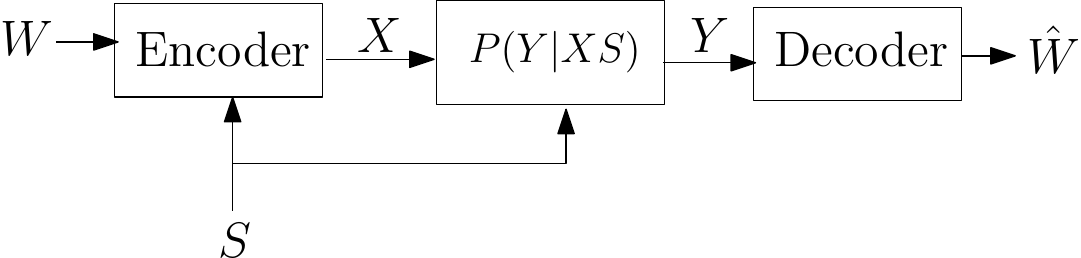}
\label{fig:gp}
}
\caption{\small{Source and Channel Coding with Side-Information}}
\label{fig:wz_gp}
\end{figure}

The basic ingredients of the constructions used to prove coding theorems for many multi-terminal problems are:
\begin{enumerate}
\item Rate-optimal \emph{point-to-point} source and channel codes,
\item Random binning,
\item Superposition coding.
\end{enumerate}
As mentioned above, it has been shown in \cite{AntonyML,AntonyFast,RVGaussianRD12} that SPARCs are rate-optimal for Gaussian channels and sources. In this paper, we show that  source and channel coding SPARCs can be combined to implement binning and superposition, thus yielding  a new class of codes for multi-terminal source and channel coding.  To illustrate how SPARCs can be used for binning, we consider the canonical examples of source coding with decoder side-information (the Wyner-Ziv problem \cite{WynerZiv}) and channel coding with encoder side-information (the Gelfand-Pinsker problem \cite{GelfandPinsker,CostaDP}). These problems are depicted in Figure \ref{fig:wz_gp}. Superposition coding using SPARCs is a natural extension of point-to-point coding and is illustrated via the Gaussian multiple-access and broadcast channels.

In Sections \ref{sec:sparc} and \ref{sec:sparc_p2p}, we review the SPARC construction and the minimum-distance performance results for source and channel coding. In Section \ref{sec:wz}, we describe how to implement random binning using SPARCs and use it to construct codes for the Wyner-Ziv problem
(Figure \ref{fig:wz}). The standard construction for this problem consists of a high-rate source codebook partitioned into bins, each of which serves as a lower-rate channel code. Due to the importance of the problem, several practical code-constructions have been proposed, e.g., \cite{Discus03,GirodCoding1, XiongCoding1,WainMart09}, but they generally fall short of the Wyner-Ziv bounds; besides they do not come with provable performance guarantees. Recently polar codes have been proposed for Wyner-Ziv coding \cite{Polarrd,Polarwzgp}. These are the first computationally efficient code constructions that are provably rate-optimal. However, these are only applicable to problems where the source and side-information distributions are discrete and symmetric.
%No low-complexity  coding schemes are known yet that achieve the Wyner-Ziv rate continuous source and side-information variables;
Elegant coding schemes such as  those based on lattices have been proposed \cite{Zamir02, ErezLZ05} for the Wyner-Ziv problem with continuous-valued source and side-information, but they have exponential encoding and decoding complexity.
%The optimal rate-distortion performance limit was obtained by Wyner and Ziv in \cite{WynerZiv}.

In Section \ref{sec:gp_multiuser}, we turn  our attention to channels with state, where the state information is known non-causally at the transmitter.  This model (Figure \ref{fig:gp})  has been studied widely in the literature \cite{GelfandPinsker, CostaDP,Zamir02,ErezSZ05}  and has found  many practical applications such as multi-antenna communication \cite{SpencerCommMag}, digital watermarking \cite{MoulinK05,ChenW01} and steganography \cite{SulliStegan}. It is the channel coding dual  of the Wyner-Ziv problem \cite{PradhanDuality,WornellDuality}.
 %shows a  channel with state known at the transmitter.
%One can think of the channel as having two inputs $A$ and $S$, of which the encoder can control $A$ but only observe $S$.
In Figure \ref{fig:gp},  the encoder knows the entire state sequence  $S$ at the beginning of communication while the decoder observes only the channel output $Y$. This capacity of this channel model was determined by Gelfand and Pinsker \cite{GelfandPinsker}. For the important special case of AWGN channels with Gaussian state, Costa \cite{CostaDP} showed that the Gelfand-Pinsker capacity is the same as the rate achievable when the decoder has full knowledge of $S$.
Since Costa's discovery of this surprising result (dubbed `writing on dirty paper'),  elegant capacity-achieving coding schemes have been developed such as  nested lattice codes \cite{Zamir02,ErezSZ05,ChenW01}, but these are generally computationally infeasible.  Several computationally efficient code designs have also been proposed, e.g., \cite{XiongDPCoding,EreztenBrink}; however they do not come with provable rate guarantees.
In Section \ref{sec:gp_multiuser}, we show how to implement Costa's coding scheme  by partitioning a high-rate SPARC channel code into bins of lower-rate source codes.

Finally in Section \ref{sec:mac_bc}, we show how to construct capacity-achieving codes for the AWGN multiple-access and broadcast channels using SPARCs.  We show that superposition codes for these channels can be implemented through a simple extension of SPARCs for point-to-point channel coding.

The analysis of SPARCs in this paper is presented with minimum-distance encoding and decoding, which is optimal but computationally inefficient. This is mainly to keep exposition simple and to highlight the main contribution of the paper -- a demonstration that binning and superposition can be easily implemented with sparse regression ensembles described by compact dictionaries. The results also hold with the feasible SPARC encoders and decoders developed in \cite{AntonyFast, KontSPARC, RVGaussianFeasible}. Further, we focus only on the achievability of the optimal information-theoretic rates and  do not discuss the SPARC error exponents obtained in \cite{AntonyML,RVGaussianRD12}. These aspects will be discussed in an extended version of this paper.

\emph{Notation}: Upper-case letters are used to denote random variables, lower-case for their realizations,  and bold-face letters to denote random vectors and matrices.  All vectors  have length $n$.
%The  source sequence  is denoted by $\bfx \triangleq (X_1, \ldots, X_n)$, and the reconstruction sequence by $\bfxh \triangleq (\hat{X}_1, \ldots, \hat{X}_n)$.
$\norm{\mathbf{X}}$ denotes the $\ell_2$-norm of vector $\mathbf{X}$, and
$\abs{\mathbf{X}} =  \norm{\mathbf{X}} / \sqrt{n}$ is the normalized version. We use natural logarithms, so entropy is measured in nats.
To limit the number of symbols introduced, we reuse notation across sections.  For example, $X$ is used to represent the channel input as well as the source; $Y$ is used to  denote both the channel output and the source side-information. The model description at the beginning of each section explains all the variables used in it.

\section{Sparse Regression Codes} \label{sec:sparc}
\begin{figure}[t]
\centering
\includegraphics[height=1.7in]{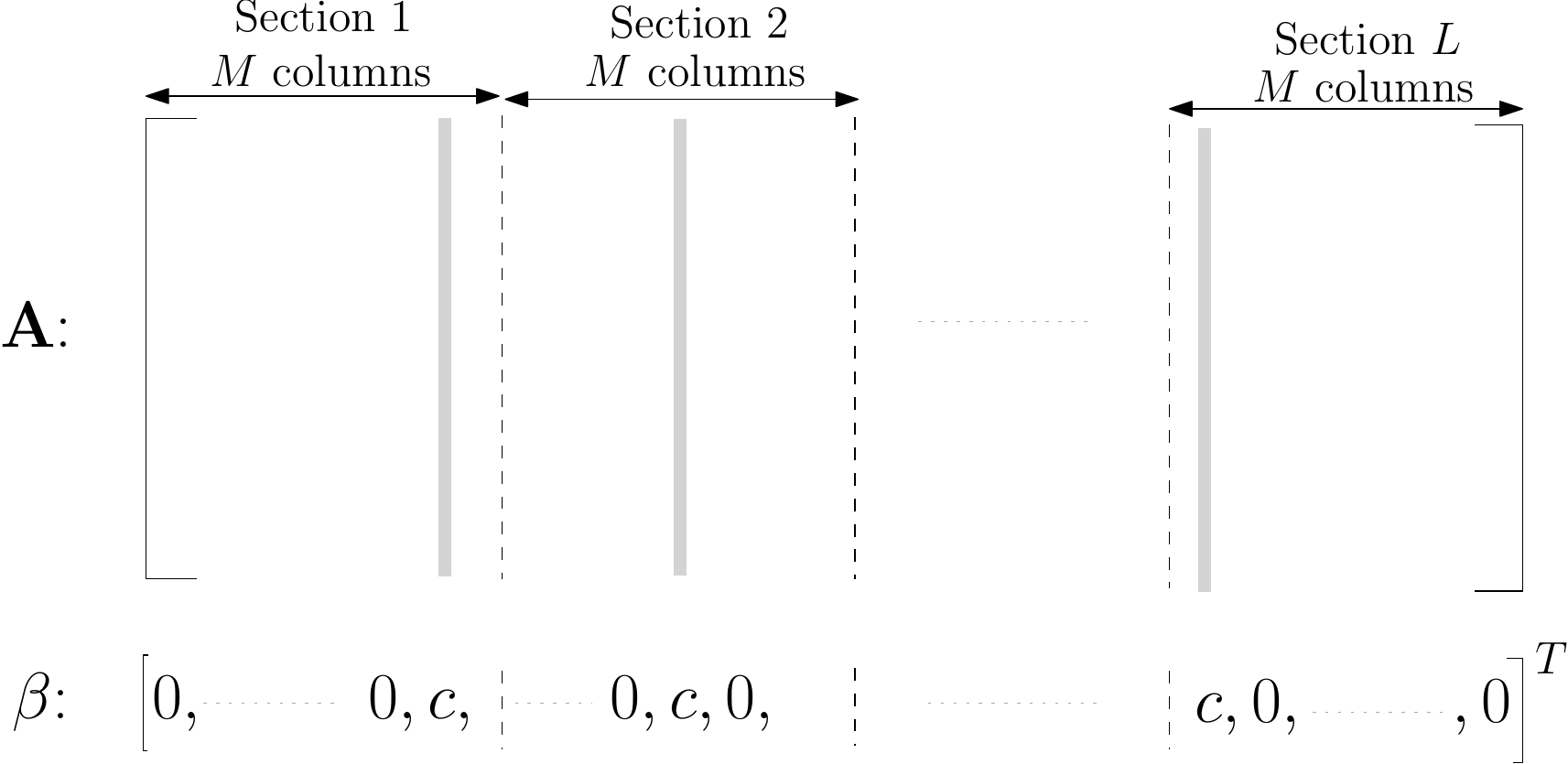}
\caption{\small{$\mathbf{A}$ is an $n \times ML$ matrix and $\beta$ is a $ML \times 1$ binary vector. The positions of the non-zeros in $\beta$  correspond to the gray columns of $\mathbf{A}$ which add to form the codeword $\mathbf{A}\beta$.}}
\vspace{-5pt}
\label{fig:sparserd}
\end{figure}

A sparse regression codebook (SPARC) is defined in terms of a design matrix $\mathbf{A}$ of dimension $n \times ML$ whose entries are i.i.d. $\mathcal{N}(0,1)$, i.e., independent zero-mean Gaussian random variables with unit variance. Here $n$ is the block length and $M$ and $L$ are integers whose values will be specified shortly in terms of $n$ and the rate $R$.  As shown in Figure \ref{fig:sparserd}, one can think of the matrix $\mathbf{A}$ as composed of $L$ sections with $M$ columns each. Each codeword is a linear combination of $L$ columns, with one column from each section.
Formally, a codeword can be expressed as  $\mathbf{A} \beta$, where $\beta$ is  an $ML \times 1$ vector $(\beta_1, \ldots, \beta_{ML})$ with the following property:  there is exactly one non-zero $\beta_i$ for  $1 \leq i \leq M$, one non-zero $\beta_i$ for $M+1 \leq i \leq 2M$, and so forth. Denote the set of all $\beta$'s that satisfy this property by $\mcb$. The non-zero values of $\beta$ are all set equal to $c \triangleq \frac{\gamma}{\sqrt{L}}$ where $\gamma$ will be specified later depending on the problem at hand.

Since there are $M$ columns in each of the $L$ sections, the total number of codewords is $M^L$. To obtain a rate of $R$ nats/sample, we therefore need
\be
M^L = e^{nR}.
\label{eq:ml_nR}
\ee
There are several choices for the pair $(M,L)$ which satisfy this. For example, $L=1$ and $M=e^{nR}$ recovers the Shannon-style random codebook; here the number of columns in the dictionary $\mathbf{A}$ is $e^{nR}$, i.e., exponential in $n$. For our constructions, we choose $M=L^b$ for some $b>1$ so that \eqref{eq:ml_nR} implies
\be   L \log L = {nR}/{b}. \label{eq:rel_nL} \ee
%Thus  $L=\Theta(\frac{n}{\log n})$, i.e., each codeword is formed by adding roughly $\frac{n}{\log n}$ i.i.d Gaussian vectors of length $n$.
 Thus $L$ is now $\Theta\left(\frac{n}{\log n}\right)$, and the number of columns $ML$ in the dictionary $\mathbf{A}$ is now $ \Theta\left(\frac{n}{\log n}\right)^{b+1}$, a \emph{polynomial}  in $n$. This reduction in dictionary complexity can be harnessed to develop computationally efficient encoders and decoders for the sparse regression code. %We note that the code structure automatically yields low decoding complexity.

Since each codeword in a SPARC is a linear combination of $L$ columns of $\mathbf{A}$ (one from each section),  codewords sharing one or more common columns in the sum will be dependent. %This will play an important role in the  analysis.
Also, SPARCs are not  linear codes since the sum of two codewords does not equal another codeword in general.

\section{SPARC for Point-to-Point Source and Channel Coding} \label{sec:sparc_p2p}
In this section, we review the performance of SPARCs for point-to-point source and channel coding under minimum distance encoding/decoding.

\subsection{Lossy Source Coding} \label{subsec:sparc_rd}
Consider an i.i.d Gaussian source $X$ with mean $0$ and variance $\sigma^2$. A rate-distortion codebook with rate $R$ and block length $n$  is a set of $e^{nR}$ length-$n$  codewords, denoted $\{\bfxh(1),\ldots, \bfxh(e^{nR}) \}$.  The quality of reconstruction is measured through the mean-squared distortion criterion
\[ d_n(\bfx, \bfxh)=\abs{\bfx -\bfxh}^2= \frac{1}{n}\sum_{i=1}^n(X_i - \hat{X}_i)^2 , \]
where $\bfxh$ is the codeword chosen to represent the source sequence $\bfx$. For this distortion criterion, an optimal encoder maps each source sequence to the codeword nearest to it in Euclidean distance.  The  rate-distortion function $R^*(D)$, the minimum rate  for which the distortion can be bounded by $D$ with high-probability, is given by  \cite{CoverThomas}
\be \label{eq:gaussian_rd} R^*(D) = \min_{p_{\hat{X}|X}: E(X-\hat{X})^2 \leq D} I(X;\hat{X}) = \frac{1}{2} \log \frac{\sigma^2}{D} \ \text{nats/sample}. \ee

For rates $R > R^*(D)$, a sparse regression codebook is defined in terms of an $n \times ML$ design matrix $\mathbf{A} \beta$, as described in the previous section. The non-zero values of $\beta \in \mcb$ are all set equal to $\sqrt{(\sigma^2-D)/L}$.
Encoding and decoding are as follows.

\emph{Minimum-distance Encoder}: This is defined by a mapping $g: \mathbb{R}^n \to \mcb$. Given the source sequence $\bfx$, the encoder determines the $\beta$ that produces the codeword closest in Euclidean distance, i.e.,
 \[ g(\bfx) = \underset{\beta \in \mcb}{\operatorname{argmin}} \ \norm{\bfx - \mathbf{A}\beta}^2.\]

\emph{Decoder}: This is a mapping $h: \mcb \to \mathbb{R}^n$. On receiving ${\beta} \in \mcb$ from the encoder, the decoder produces reconstruction $h(\beta) = \mathbf{A}\beta$. %\textbf{Decoder complexity low}

The probability of error at distortion-level $D$ of a rate-distortion code $\mathcal{C}_n$ with block length $n$ and encoder and decoder mappings $g,h$  is
\be P_{e}(\mathcal{C}_n, D) = P\left(\abs{\bfx - h(g(\bfx))}^2 > D\right).  \ee

It was shown in \cite{RVGaussianRD12} that SPARCs can achieve the optimal rate-distortion function with the optimal error-exponents for i.i.d Gaussian sources for all distortions $D$ such that $D/\sigma^2 < x^*$, where $x^* \approx 0.2032$ is the solution of the equation
\be 1 + \frac{1}{2} \log x =x. \label{eq:xstar_def}\ee

\begin{fact} \cite{RVGaussianRD12}
For $D \in (0, \sigma^2)$, let $R_{sp}(D) = \max\{ \frac{1}{2} \log \frac{\sigma^2}{D}, \ 1-\frac{D}{\sigma^2} \}$.  Fix  rate $R > R_{sp}(D)$, $\epsilon>0$ and  $b>b_1$ where
\be b_1 = \frac{2.5 R}{R- 1  + D/\sigma^2}. \ee
For all $n$, let $\mathcal{C}_n$ be a rate $R$ SPARC defined by an $n  \times L_n M_n$ design matrix with i.i.d $\mc{N}(0,1)$ entries, where $L_n$ is determined by \eqref{eq:rel_nL} and $M_n =L_n^b$. Then for all sufficiently large $n$, $P_{e}(\mathcal{C}_n, D) < \epsilon$.
\label{fact:rd_sparc}
\end{fact}

%\textbf{ \cite{RVGaussianRD12} also describes error exponent performance, but for clarity  we do not present it here. they can be easily included.}

 Fact \ref{fact:rd_sparc} implies that SPARCs achieve the optimal rate-distortion function  for $0 < \frac{D}{\sigma^2}  < x^*$ where $x^* \approx 0.2032$ is the solution of \eqref{eq:xstar_def}. For $x^*\leq \frac{D}{\sigma^2} \leq 1$, the minimum achievable rate of Fact \ref{fact:rd_sparc} $(1-\frac{D}{\sigma^2})$  is larger than the optimal rate-distortion function.
 %In this region $R_{sp}(D)$ can also be achieved by time-sharing between the points $\frac{D}{\sigma^2}=x^*$ and $\frac{D}{\sigma^2}=1$.
%The rate-distortion performance is shown in Figure \ref{fig:sparc_perf}.

\subsection{Communication over an AWGN Channel} \label{subsec:ch_cod}
Consider an AWGN channel with input $X$ and output $Y$ defined by
\[
Y = X + Z
\]
where $Z \sim \mathcal{N}(0, N)$ is a noise variable independent of $X$. There is an average power constraint $P$ on the input $X$. Denote by $v$ the signal-to-noise ratio ${P}/{N}$. It was shown in \cite{AntonyML} that SPARCs can achieve the capacity $\frac{1}{2} \log (1+v)$ with the probability of error decaying exponentially with $n$.

\emph{Encoder}: This is a mapping $g: \mcb \to \mathbb{R}^n$. Each message in the set $\{1,\ldots, M^L=e^{nR} \}$ is indexed by a unique $\beta \in \mcb$. The non-zero values of $\beta$ are all equal to $\sqrt{P/L}$. To transmit the message corresponding to $\beta$,  the encoder produces the channel input  $\bfx = \mathbf{A}\beta$. %\textbf{Encoder complexity low}

\emph{Minimum-distance Decoder}: This is defined by a mapping $h: \mathbb{R}^n \to \mcb$. Upon receiving the output sequence $\bfy$, the encoder determines the $\beta$ that produces the codeword closest in Euclidean distance, i.e.,
 \[ \hat{\beta} = h(\bfy) = \underset{\beta \in \mcb}{\operatorname{argmin}} \ \norm{\bfy - \mathbf{A}\beta}^2.\]

 The average probability of error of a code $\mathcal{C}_n$ with block length $n$ and encoder and decoder mappings $g,h$  is
\be P_{e}(\mathcal{C}_n) = \frac{1}{M^L} \sum_{\beta \in \mcb} P\left( \hat{\beta} \neq \beta \mid \mathbf{X} = \mathbf{A} \beta  \right).  \ee

The performance of SPARC for channel coding is given below.

Let $v^* \approx 15.8$ be the solution to $(1+v^*)\log(1+v^*)=3v^*$. Define
\be
b_0 (v) =  \left\{
\begin{array}{ll}
 \frac{4v (1+v) \log(1+v)}{[(1+v)\log(1+v) -v]^2} & \quad v < v^* \\
\\
 \frac{(1+v) \log(1+v)}{(1+v)\log(1+v) -2v} & \quad v \geq v^*
\end{array}
\right.
\label{eq:b0_def}
\ee
We note that $b_0$ asymptotically approaches $1$ with growing $v$.
\begin{fact} \cite{AntonyML}
Fix rate $R < C = \frac{1}{2}\log(1+v)$, $b>b_0(v)$ and $\epsilon>0$.
 For all $n$, let $\mathcal{C}_n$ be a rate $R$ SPARC defined by an $n  \times L_n M_n$ design matrix with i.i.d $\mc{N}(0,1)$ entries, where $L_n$ is determined by \eqref{eq:rel_nL} and $M_n =L_n^b$. Then for all sufficiently large $n$, $P_{e}(\mathcal{C}_n) < \epsilon$.
\label{fact:ch_coding}
\end{fact}

\section{SPARC for Lossy Compression with Decoder Side-Information} \label{sec:wz}

 In this  section, we construct SPARCs to achieve the optimal Wyner-Ziv rate for Gaussian sources.
Consider an i.i.d Gaussian source $X \sim \mc{N}(0, \sigma^2)$ to be compressed with mean-squared distortion $D$. The decoder side-information $Y$ is noisy version of $X$  and is related to $X$ by $Y=X + Z$, where $Z \sim \mc{N}(0,N)$ is independent of $X$. The sequence $\mathbf{Y}$ is available at the decoder non-causally.  If $\mathbf{Y}$ were available at the encoder as well, the optimal strategy is to compress $\mathbf{Z}= \mathbf{Y} - \mathbf{X}$ to within distortion $D$; the minimum rate required for this is $\frac{1}{2} \log \frac{\text{Var}(X|Y)}{D}$ nats/sample. Wyner and Ziv showed in \cite{WynerZiv} that this rate is achievable even when $\mathbf{Y}$ is available at only the decoder.

\begin{figure*}[t]
\centering
\includegraphics[height=2in]{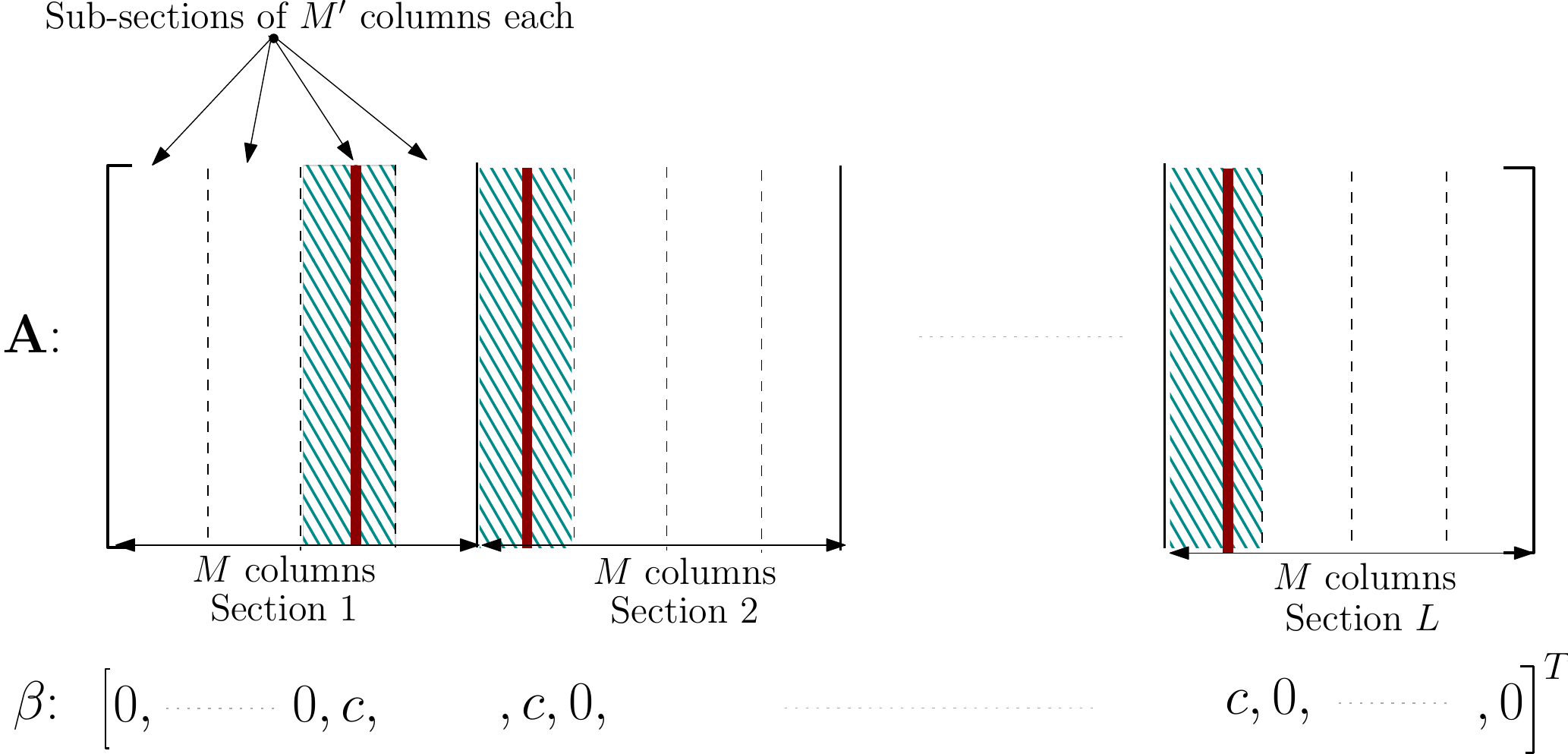}
\caption{\small{Each section is divided into subsections of $M'$ columns. A bin is formed by specifying a subsection in each of the $L$ sections as shown by the shaded regions.}}
\label{fig:binning}
\end{figure*}

Before presenting the SPARC construction, we briefly review the main ideas in the Wyner-Ziv random coding scheme\cite{WynerZiv}. Define an auxiliary random variable $U$  jointly distributed with $X$ according to
\be \label{eq:uxv} U = X  + V \ee where $V \sim \mc{N}(0, Q)$ is independent of $X$.  The idea is that the decoder first recovers $U$, and then produces $\hat{X}$ as the best estimate of $X$ given $U$ and $Y$.
 The codebook consists of length-$n$ vectors  chosen i.i.d according to the marginal distribution of $U$. The encoder
 %`quantizes' the source vector $\bfx$ to this codebook, i.e., it
 attempts to find a codeword $\mathbf{U}$  whose empirical joint distribution with $\bfx$ is close to \eqref{eq:uxv}. From the rate-distortion theorem, this step will be successful if the codebook size is {at least} slightly larger than $e^{nI(U;X)}$.
 Since the decoder has $\bfy$, the index of the chosen codeword $\mathbf{U}$ is not sent in its entirety; instead we divide the codebook into $e^{nR}$ equal-sized bins and send only the index of the bin containing the codeword. Thus the information rate to the decoder is  $R$ nats/source sample which is less than the rate $I(U;X)$ required to convey the precise codeword index.

 The decoder's task is to recover the codeword $\mathbf{U}$ using the bin index and the side-information $\bfy$. This is equivalent to a \emph{channel decoding} problem.
%since we can think of $Y$ as being obtained by passing $U$ through an AWGN channel with appropriate `noise' variance.
We can correctly distinguish $\mathbf{U}$ from the other codewords in the bin if number of codewords in each bin is exponentially less than $e^{nI(U;Y)}$. Combining this with the minimum codebook size for quantization, we see that the number of bins $e^{nR}$ should satisfy
\[ e^{nR} > \frac{e^{nI(U;X)}}{e^{nI(U;Y)}} \] { or } \[ R > I(U;X) - I(U;Y) = \frac{1}{2} \log \frac{\text{Var}(X|Y)}{D} \]
 where the last inequality is obtained by setting $Q= \frac{\text{Var}(X|Y) D}{\text{Var}(X|Y) - D}$. After decoding $\mathbf{U}$, the decoder reconstructs $\mathbf{\hat{X}}$ as the MMSE estimate of $\bfx$ given $(\mathbf{U},\mathbf{Y})$. It can be verified that expected squared-error distortion is $D$.

%The essence of the Wyner-Ziv coding scheme is a good quantization codebook partitioned into bins, each of which is a good channel code.
We now show that the above coding scheme with binning can be implemented with SPARCs.
%First construct a SPARC codebook to perform the quantization from $X$ to $U$.
The relation \eqref{eq:uxv} can be equivalently written in terms of the reverse test channel as
\be
X= a U + V'
\ee
where $a = \frac{\sigma^2}{\sigma^2+Q}$, and $V' \sim \mathcal{N}(0,\frac{\sigma^2 Q}{\sigma^2 + Q}) $ is independent of $U$. The first step of the coding scheme is equivalent to quantizing the source sequence $\bfx$ to a codeword $a\mathbf{U}$ with mean-squared distortion at most $\frac{\sigma^2 Q}{\sigma^2+Q}$. We can use a SPARC to perform this quantization by choosing a design matrix with parameters satisfying the specifications in Fact \ref{fact:rd_sparc}.

Instead of sending the codeword index $\beta$ to the decoder in its entirety, we divide each section of the design matrix $\mathbf{A}$ into subsections of $M'$ columns each as shown in Figure \ref{fig:binning},  and only send information to indicate which subsection in each of the $L$ sections of $\beta$ contains a non-zero.  More precisely, we send the decoder a tuple $(p_1,\ldots,p_L)$ where $p_i \in \{1, \ldots, \frac{M}{M'} \}$  indicates a subsection in the $i$th section of $\mathbf{A}$.
This strategy is equivalent to binning: a bin is now a subset of the codebook  consisting of  codewords corresponding to $\beta$'s  with ones in the  sections specified by $(p_1,\ldots,p_L)$.  The codebook is thus divided into $\left(\frac{M}{M'}\right)^L$ bins and  the rate $R$ required to send  the bin index to the decoder is determined as
\be e^{nR} =  \left({M}/{M'}\right)^L   \label{eq:rd_rmm'}. \ee  We note that each bin is itself a smaller sparse superposition codebook with $M'^L$ codewords, defined by a $n \times M'L$ sub-matrix of $\mathbf{A}$.

The decoder side-information variable $Y$ is related to $U$ as
\be
Y= X+ Z = aU + V' +Z
\ee
where $U,V'$ and $Z$ are mutually independent. The problem of recovering $U$ from $Y$ at the decoder is a channel decoding problem over a channel with signal-to-noise ratio given by
\be \textsf{snr}= \frac{a^2 \text{Var}(U)} {\text{Var}(V') + \text{Var}(Z)} %= \frac{P^2/(P+Q)}{N + PQ/(P+Q)}
=  \frac{\sigma^4}{\sigma^2 Q + (\sigma^2+Q)N}. \label{eq:snr_def}\ee
Since Fact \ref{fact:ch_coding} shows shown that SPARCs can achieve the AWGN channel capacity, the decoder can perfectly recover $\mathbf{U}$ if the number of codewords in each bin  satisfies
\[ M'^L < \exp\left(n \frac{1}{2} \log(1 + \textsf{snr})\right).\]

The above SPARC coding scheme and its performance are formalized below.
\begin{defi}[Nested Sparse Regression Codebook] A nested sparse regression codebook with rates $(R_1, R_2)$ and block length $n$ is defined by an
$n \times ML$ design matrix $\mathbf{A}$ with i.i.d $\mc{N}(0,1)$ entries, where $M^L =e^{nR_1}$. Each section of $M$ columns is divided into sub-sections of $M'$ columns each, where $M'$ is determined by
$M'^L =e^{nR_2}$. The codebook consists of codewords $\mathbf{A}\beta$, where $\beta \in \mcb$ contains one non-zero element in each of the $L$ sections.
%Further, if $M=L^b$ and $M'=L^{b'}$ for some constants $b,b' \geq 1$, the nested SPARC is parameterized by the tuple $(n, R_1, R_2, b, b')$.
\end{defi}

\begin{thm}
Fix $R_1 > \max\{ \frac{1}{2} \log \frac{\sigma^2 + Q}{Q}, \frac{\sigma^2}{\sigma^2 + Q} \}$ and $R_2 < \frac{1}{2} \log (1+ \textsf{snr})$ where
\[ Q  = \frac{\text{Var}(X|Y) D}{\text{Var}(X|Y) - D} \]
and $\textsf{snr}$ is given by \eqref{eq:snr_def}. Then for any $\epsilon >0$ and all sufficiently large $n$,  there exists a rate $R_1-R_2$  code $\mc{C}_n$ with $P_e(\mc{C}_n,D) < \epsilon$ where $\mc{C}_n$ is defined by a nested sparse regression codebook with rates $(R_1,R_2)$ whose $n \times ML$ design matrix satisfies the following:  $M=L^b$ where
\[ b > \max \left\{ \frac{2.5 R_1}{R_1 - \sigma^2/(\sigma^2 +Q)}, \ \frac{R_1}{R_2} b_0(\textsf{snr}) \right\} \]
and $L$ is determined by  $b L \log L = n R_1$.  (The function $b_0(.)$ is defined in \eqref{eq:b0_def}).
\label{thm:wz}
\end{thm}

Define
\begin{equation*}
D^* = \frac{x^* \sigma^2}{1 + x^* \sigma^2/ N}
\end{equation*}
where $x^* \approx 0.2032$ is the solution of \eqref{eq:xstar_def}.
\begin{corr}
For $D \in (0, D^*)$, sparse regression codes achieve the optimal Wyner-Ziv rate-distortion function for Gaussian sources given by
$\frac{1}{2} \log \frac{\text{Var}(X|Y)}{D}$.
\end{corr}
\begin{IEEEproof}
It can be verified that for $D \in (0, D^*)$, the lower bound on $R_1$  specified by the theorem becomes $ \frac{1}{2} \log \frac{\sigma^2 + Q}{Q}$. The corollary then follows by choosing $R_1 = \frac{1}{2} \log \frac{\sigma^2 + Q}{Q} + \epsilon$ and
$R_2 = \frac{1}{2} \log (1+ \textsf{snr}) - \epsilon$. This yields an achievable rate $R_1-R_2= \frac{1}{2} \log \frac{\text{Var}(X|Y)}{D} + 2 \epsilon$
where $\epsilon >0$ can be arbitrarily small.
\end{IEEEproof}

\emph{Proof of Theorem \ref{thm:wz}}:

Fix block length $n$ and rates $R_1, R_2$. Choose a  $n \times ML$ design matrix $\mathbf{A}$ with $M=L^b$ and $bL \log L = nR_1$ where $b$ is greater than the minimum value specified by the theorem. Each section of $\mathbf{A}$ is partitioned into sub-sections of  $M'$ columns each, where $M'^L = e^{nR_2}$.

The $U$-codebook consists of all vectors $\mathbf{A}\beta$ such that $\beta \in \mcb$ and the non-zero entries in $\beta$ are all equal to
$\sqrt{\frac{\sigma^2 + Q}{L}}$.
Let  \[ a= \frac{\sigma^2}{\sigma^2 + Q}. \]

\emph{Encoder}: Given source sequence $\mathbf{X}$, find the codeword $\mathbf{U}$ from the SPARC such that $a\mathbf{U}$ is closest to $\mathbf{X}$ in Euclidean distance. Specifically, determine
\[ \beta^* = \underset{\beta \in \mcb}{\operatorname{argmin}} \ \norm{\bfx - a\mathbf{A}\beta}^2. \]
Send the decoder a tuple $(p_1,\ldots,p_L)$ where $p_i \in \{1, \ldots, \frac{M}{M'} \}$  indicates the subsection in the $i$th section of $\mathbf{A}$ where
$\beta^*$ contains a non-zero element.
The rate  to the decoder is
\[ \frac{1}{n} \log \left(\frac{M}{M'}\right)^L = R_1-R_2. \]

\emph{Decoder}: The decoder first determines the $n \times M'L$ sub-matrix corresponding to the subsections of $\mathbf{A}$ indicated by $p_1, \ldots, p_L$. We denote this sub-matrix $\mathbf{A}_{bin}$. $\mathbf{A}_{bin}$ defines a SPARC $\mathbf{A}_{bin}\beta$ where $\beta \in \mathcal{B}_{M',L}$ contains one non-zero value equal to $\sqrt{(\sigma^2 + Q)/{L}}$ in each of its $L$ sections. The decoder now reconstructs $\mathbf{\hat{U}} = \mathbf{A}_{bin}\hat{\beta}$ where
\be
\hat{\beta} = \underset{\beta \in \mc{B}_{M',L}}{\operatorname{argmin}} \ \norm{\bfy - a \mathbf{A}_{bin}\beta}^2.
\ee
Finally, the source sequence is reconstructed as
\be \bfxh= \left(\frac{1}{Q} + \frac{1}{\sigma^2} + \frac{1}{N} \right)^{-1} \left( \frac{\mathbf{\hat{U}}}{Q} + \frac{\mathbf{Y}}{N}  \right) \ee

\emph{Error Analysis}:
Let $\delta >0$ be such that
\be
\begin{split}
R_1  > \max\left\{ \frac{1}{2} \log \frac{(\sigma^2 + Q) (1 + \delta)}{Q}, 1 - \frac{Q}{(\sigma^2 + Q) (1 + \delta)} \right\}
\end{split}
\ee
%where $\textsf{snr}_{\sigma^2-\delta}$ is defined by replacing $\sigma^2$ in \eqref{eq:snr_def} with $\sigma^2 - \delta$.
The probability of the error event $\mc{E}$, can be decomposed as $P(\mc{E}) = P(\mc{E}_1 \cup \mc{E}_2 \cup \mc{E}_3)$
where
%\[ P(\mc{E}_1) = P(\abs{\abs{X}^2 - \sigma^2}^2 > \delta), \]
$\mc{E}_1$ is the event that $\abs{\bfx}^2 > \sigma^2(1 + \delta)$,
$\mc{E}_2$ is the event of error at the encoder, and $\mc{E}_3$ the event of error at the decoder. We have
\be
P(\mc{E}_1) = P\left(\abs{\bfx}^2 > \sigma^2(1 + \delta) \right) < \frac{\epsilon}{3}
\label{eq:wz_error1}
\ee
for sufficiently large $n$ from standard results on large-deviations \cite{DemboZbook}. Next, we have
\be
P(\mc{E}_2 \mid \mc{E}_1^c) = P\left(\min_{\beta \in \mc{B}_{M,L}} \abs{\bfx - a\mathbf{A}\beta}^2 > \frac{\sigma^2 Q}{\sigma^2 + Q}\right) < \frac{\epsilon}{3}
\label{eq:wz_error2} \ee
for sufficiently large $n$. This follows from Fact \ref{fact:rd_sparc} since $R_1$ and $b$ satisfy the conditions specified in Fact \ref{fact:rd_sparc} for compressing source sequences of variance up to $\sigma^2(1+\delta)$ at  distortion-level $\frac{\sigma^2 Q}{\sigma^2 + Q}$.
Finally, we bound
\[ P(\mc{E}_2 \mid \mc{E}_1^c, \mc{E}_2^c ) = P\left(\underset{\beta \in \mc{B}_{M',L}}{\operatorname{argmin}} \ \norm{\bfy - a \mathbf{A}_{bin}\beta}^2 \neq \beta^*\right). \]
Let the number of columns in each sub-section $M'=L^{b'}$. Using $M'^L = e^{nR_2}$ we have
\be
b' = \frac{n R_2}{L \log L} = b \frac{R_2}{R_1} > b_0 (\textsf{snr})
\ee
where the last inequality is due to the minimum value of $b$ specified by the theorem. Since
\[ R_2 <  \frac{1}{2} \log (1 + \textsf{snr}) \]
and $b'> b_0(\textsf{snr})$, the $n \times M'L$ design matrix $\mathbf{A}_{bin}$ satisfies the conditions of Fact \ref{fact:ch_coding} for signal-to-noise ratio given by \eqref{eq:snr_def}. Hence for sufficiently large $n$, $P(\mc{E}_2 \mid \mc{E}_1^c, \mc{E}_2^c ) < \epsilon/3$. Combining this with  \eqref{eq:wz_error1} and  \eqref{eq:wz_error2}, we have $P(\mc{E}) < \epsilon$. % for sufficiently large $n$.
\hfill\ensuremath{\blacksquare}

\section{SPARC for writing on Dirty Paper} \label{sec:gp_multiuser}

The AWGN channel with state is defined by the relation  $Y= X + S + Z$,
where the state $S \sim \mc{N}(0, \sigma^2_s)$ is independent of the additive noise $Z \sim \mc{N}(0,N)$. There is an average power constraint $P$ on the input sequence $\bfx$. The state sequence $\bfs \sim$ i.i.d $\mc{N}(0, \sigma^2_s)$ is known non-causally at the encoder.

 We first review the main ideas behind Costa's capacity-achieving coding scheme \cite{CostaDP} for this channel.
 %If the decoder knew the state sequence $\bfs$, it could just subtract it from the output and a rate of $\frac{1}{2} \log_2 \left(1 + \frac{P}{N} \right)$  is achievable.
 The state sequence $\bfs$ (known only at the encoder) is used in two ways: part of it is used for coding and the rest is treated as noise.
Define an auxiliary random variable $U$ as
\be U= X  + \alpha S  \label{eq:u_gp}\ee
where $X \sim \mc(0,P)$ is independent of $S$ and $\alpha \in (0,1)$ is a constant specified later. The channel codebook consists of $e^{nR_1}$ $U$-sequences chosen i.i.d
$\mc{N}(0, P+ \alpha^2 \sigma^2_s)$.  We divide this codebook into $e^{nR}$ equal-sized {bins} with each bin representing a message.
To transmit message $m \in \{1, \ldots, e^{nR} \}$, the encoder observes the state sequence $\bfs$ and attempts to find a codeword $\mathbf{U}$ {within} bin $m$
whose empirical joint distribution with $\bfs$ is close to \eqref{eq:u_gp}.
%This step can be thought of as quantizing the state sequence $\mathbf{S}$ to a codeword $\mathbf{U}$.
From rate-distortion theory, this step will be successful if the number of sequences in each bin  $e^{n(R_1-R)}$ is  larger than $e^{n I(U;S)}$.  The encoder then forms the channel input sequence $\bfx$ as $\mathbf{U}-\alpha \bfs$.
%so that the empirical joint distribution of $(\mathbf{U},\bfx,\bfs)$ is  well-approximated by \eqref{eq:u_gp}.

The channel receives the output sequence $\bfy$ according to
\be Y = X+S+Z = U + (1-\alpha) S + Z  \label{eq:ch_out}\ee
and attempts to decode $\mathbf{U}$. This is effectively an AWGN channel  decoding operation, which will be successful if $R_1 < I(U;Y)$. Combining this with the lower bound $R_1-R > I(U;S)$, we see that any rate $R < I(U;Y) - I(U;S)$
 is achievable. The right-side of the inequality is equal to the channel capacity $\frac{1}{2} \log (1 + {P}/{N})$ for the joint distribution given by \eqref{eq:u_gp} and \eqref{eq:ch_out} for $\alpha= \frac{P}{P+ N}$.

We now show how to implement the above coding scheme with a nested SPARC. Define a nested SPARC with rates $(R_1,R_1-R)$ through an $n \times ML$ matrix $\mathbf{A}$ with $M^L = e^{nR_1}$. As in Section \ref{sec:wz}, each bin corresponds to a SPARC defined by a sub-matrix of $\mathbf{A}$,  consisting of $L$ subsections of $M'$ columns. We note that $M'^L= e^{n (R_1-R)}$ which implies that the number of bins is $(M/M')^L = e^{nR}$.  Thus each message indexes a unique bin of the nested SPARC or equivalently, a unique sub-matrix of $\mathbf{A}$ .

The relation \eqref{eq:u_gp} can be be equivalently written in terms of the reverse test channel as
\be
S= \kappa U + X'
\label{eq:sux'}
\ee
where $\kappa = {\alpha \sigma_s^2}/(P+ \alpha^2 \sigma_s^2)$ and $X' \sim \mc{N}(0, \frac{P \sigma_s^2}{P + \alpha^2\sigma_s^2})$ is independent of
$U$. Given the message and state sequence $\bfs$, the encoder needs to quantize the state sequence $\bfs$ to a codeword $\kappa \mathbf{U}$ (within the bin indexed by the message) with mean-squared distortion at most $\text{Var}(X')$. The SPARC defined by the message bin can reliably perform this quantization if the corresponding sub-matrix of $\mathbf{A}$ has parameters satisfying the specifications in Fact \ref{fact:rd_sparc}. Using \eqref{eq:sux'},  the channel law \eqref{eq:ch_out} can be written as
\be
Y = U + (1-\alpha) S + Z = U + (1-\alpha)\kappa U + (1-\alpha) X' + Z
\ee
where $U, X', Z$ are mutually independent. Thus the decoder has to recover the codeword $\mathbf{U}$ transmitted over a channel with effective signal-to-noise ratio
{\small{
\be
\textsf{snr} = \frac{ (1+ (1-\alpha)\kappa)^2 \text{Var}(U)}{(1-\alpha)^2 \text{Var}(X') + N}  =
\frac{ (1+ (1-\alpha)\kappa)^2  (P + \alpha^2 \sigma_s^2)}{\frac{(1-\alpha)^2 P \sigma_s^2}{P + \alpha^2\sigma_s^2}+N}.
\label{eq:chsnr}
\ee
}}
For all $R_1 < \frac{1}{2} \log (1 + \textsf{snr})$, this step is successful with high probability if the design matrix $\mathbf{A}$ satisfies the specifications in Fact \ref{fact:ch_coding}.

The performance of this coding scheme is formalized in the following theorem.

\begin{thm}
  Fix $\alpha \in (0,1)$. Let $\kappa =\frac{\alpha \sigma_s^2}{P+ \alpha^2 \sigma_s^2}$ and $\textsf{snr}$ be defined by \eqref{eq:chsnr}. Fix $R_1 < \frac{1}{2}\log ( 1 + \textsf{snr})$ and $R_2 > \max \{ \frac{1}{2}\log\left(1 +  \frac{\alpha^2 \sigma_s^2}{P} \right), \frac{\alpha^2 \sigma_s^2}{P+ \alpha^2 \sigma_s^2} \}$ such that $R_1 > R_2$. There exists a rate $R_1-R_2$  code $\mc{C}_n$ with $P_e(\mc{C}_n) < \epsilon$ where $\mc{C}_n$ is defined by a nested sparse regression codebook with rates $(R_1,R_2)$ whose $n \times ML$ design matrix satisfies the following:  $M=L^b$ where
\[ b > \max \left\{ \frac{2.5 R_1}{R_2 - {\alpha^2 \sigma_s^2}/(P+ \alpha^2 \sigma_s^2)}, \  b_0(\textsf{snr}) \right\} \]
and $L$ is determined by  $b L \log L = n R_1$.  (The function $b_0(.)$ is defined in \eqref{eq:b0_def}).
\label{thm:gp}
\end{thm}

\begin{corr}
 Let $x^* \approx 0.2032$ be the solution of the equation \eqref{eq:xstar_def}.
For $\frac{P \sigma_s^2}{(P+N)^2} \geq \frac{1}{x^*} - 1$, sparse regression codes achieve the channel capacity
$\frac{1}{2} \log \left(1 + \frac{P}{N}\right)$.
\end{corr}
\begin{IEEEproof}
It can be verified that when we choose $\alpha = P/(P+N)$ and  $P,N, \sigma_s^2$ satisfy the above condition, the lower bound on $R_2$ specified by the theorem becomes
$ \frac{1}{2}\log\left(1 +  \frac{\alpha^2 \sigma_s^2}{P} \right)$. The corollary then follows by choosing
$R_1 = \frac{1}{2} \log (1+ \textsf{snr}) - \epsilon$ and
$R_2 = \frac{1}{2}\log\left(1 +  \frac{\alpha^2 \sigma_s^2}{P} \right) +  \epsilon$. This yields an achievable rate
$R_1-R_2= \frac{1}{2} \log (1 + P/N) - 2 \epsilon$ where $\epsilon >0$ can be arbitrarily small.
\end{IEEEproof}

\emph{Proof of Theorem \ref{thm:gp}}:

Fix block length $n$ and rates $R_1, R_2$. Choose a  $n \times ML$ design matrix $\mathbf{A}$ with $M=L^b$ and $b$  greater than the minimum value specified by the theorem. The $U$-codebook consists of all vectors $\mathbf{A}\beta$ such that $\beta \in \mcb$ and the non-zero entries in $\beta$ are all equal to
$\sqrt{(P + \alpha^2 \sigma_s^2 )/L}$.

We have $M^L = e^{nR_1}$, and each section of $\mathbf{A}$ is partitioned into sub-sections of  $M'$ columns each where $M'^L= e^{nR_2}$.
Each of the $e^{n(R_1 - R_2)}$ messages corresponds to a unique tuple $(p_1, \ldots, p_L)$ where $p_i \in  \{1, \ldots, \frac{M}{M'} \}$.

\emph{Encoder}: The message $(p_1, \ldots, p_L)$ indexes an $n \times M'L$ sub-matrix of $\mathbf{A}$. This sub-matrix denoted by $\mathbf{A}_{bin}$.  Find the codeword $\mathbf{U}$ from the SPARC $\mathbf{A}_{bin}$ such that $ \kappa \mathbf{U}$ is closest to $\bfs$ in Euclidean distance. Specifically, determine
\[ \beta^* = \underset{\beta \in \mc{B}_{M',L}}{\operatorname{argmin}} \ \norm{\bfs - \kappa\mathbf{A}_{bin}\beta}^2 \]
and transmit
\[ \bfx = \mathbf{A}_{bin}\beta^* - \alpha \bfs.\]

\emph{Decoder}: Given channel output $\mathbf{Y}$, find the codeword $\mathbf{U}$ from the SPARC such that $ (1 + (1-\alpha) \kappa)\mathbf{U}$ is closest to $\bfy$ in Euclidean distance. Specifically, determine
\[ \hat{\beta} = \underset{\beta \in \mcb}{\operatorname{argmin}} \ \norm{\bfy - (1 + (1-\alpha) \kappa) \mathbf{A}\beta}^2. \]
Decode the message as $(\hat{p}_1,\ldots,\hat{p}_L)$ where $\hat{p}_i \in \{1, \ldots, \frac{M}{M'} \}$  indicates the subsection in the $i$th section of $\mathbf{A}$ where $\hat{\beta}$ contains a non-zero element.

\emph{Error Analysis}:
Let $\delta >0$ be such that
{\small{
\be
\begin{split}
R_2  > \max\left\{ \frac{1}{2} \log \left((1+\delta) \frac{(P + \alpha^2 \sigma_s^2)}{P} \right), 1 - \frac{P}{(P + \alpha^2 \sigma_s^2 ) (1 + \delta)} \right\}
\end{split}
\ee
}}
The probability of error can be decomposed as $P(\mc{E}) = P(\mc{E}_1 \cup \mc{E}_2 \cup \mc{E}_3)$
where
%\[ P(\mc{E}_1) = P(\abs{\abs{X}^2 - \sigma^2}^2 > \delta), \]
$\mc{E}_1$ is the event that $\abs{\bfs}^2 > \sigma_s^2(1 + \delta)$,
$\mc{E}_2$ is the event of error at the encoder, and $\mc{E}_3$ the event of error at the decoder. We have
\be
P(\mc{E}_1) = P\left(\abs{\bfs}^2 > \sigma_s^2(1 + \delta) \right) < \frac{\epsilon}{3}
\label{eq:gp_error1}
\ee
for sufficiently large $n$.  Letting $M' = L^{b'}$, we have
\be
b' = \frac{n R_2}{L \log L} = b \frac{R_2}{R_1} > \frac{2.5 R_2}{R_2 - {\alpha^2 \sigma_s^2}/(P+ \alpha^2 \sigma_s^2)}
\ee
where the last inequality is due to the minimum value of $b$ specified by the theorem.
We then have
\be
P(\mc{E}_2 \mid \mc{E}_1^c) = P\left(\min_{\beta \in \mc{B}_{M',L}} \abs{\bfs - \kappa\mathbf{A}_{bin}\beta}^2
> \frac{P\sigma_s^2}{P + \alpha^2 \sigma_s^2}\right) < \frac{\epsilon}{3}
\label{eq:gp_error2} \ee
for sufficiently large $n$. This follows from Fact \ref{fact:rd_sparc} since $R_2$ and $b'$ satisfy the conditions specified in Fact \ref{fact:rd_sparc} for compressing sequences $\bfs$ of variance up to $\sigma_s^2(1+\delta)$ at  distortion-level $\frac{P\sigma_s^2}{P + \alpha^2 \sigma_s^2}$.
Finally, $P(\mc{E}_2 \mid \mc{E}_1^c, \mc{E}_2^c )$ is given by
\be \begin{split}   P\left(\underset{\beta \in \mc{B}_{M,L}}{\operatorname{argmin}}  \  \norm{\bfy - (1 + (1-\alpha) \kappa) \mathbf{A}\beta}^2 \neq \beta^*\right)
 < \epsilon/3  \end{split}\label{eq:gp_error3}\ee
 from Fact \ref{fact:ch_coding} since $R_1< \frac{1}{2} \log (1 + \textsf{snr})$ and $b> b_0(\textsf{snr})$ for $\textsf{snr}$ given by \eqref{eq:chsnr}. Combining \eqref{eq:gp_error1}, \eqref{eq:gp_error2} and \eqref{eq:gp_error3}, we conclude that $P(\mc{E}) < \epsilon$.
\hfill\ensuremath{\blacksquare}

\section{SPARC for Gaussian Multiuser Channels} \label{sec:mac_bc}
\subsection{The AWGN Multiple-Access Channel}
In a multiple-access channel (MAC), several users simultaneously transmit to a single receiver. For simplicity let us consider the case of two users, each with average power constraint $P$, transmitting information at rates $R_1$ and $R_2$, respectively. The receiver of the AWGN MAC observes the output
\[ Y= X_1 + X_2 + Z \] where $X_1,X_2$ denote the channel inputs of the two transmitters and $Z \sim \mc{N}(0, N)$ is the channel noise independent of  $X_1$ and $X_2$. The capacity region for this channel is well-known \cite{CoverThomas}  and is given by
\begin{equation}
\begin{split}
R_1 < \frac{1}{2} \log (1 + \frac{P}{N}), & \quad
R_2 <  \frac{1}{2} \log (1 + \frac{P}{N}),  \\
R_1+ R_2  & <  \frac{1}{2} \log(1 + \frac{2P}{N}). \label{eq:mac_r1r2}
\end{split}
\end{equation}
We now show how to achieve the corner points of this rate region using SPARCs. The remaining rate points in the region can be achieved through time-sharing.
The key observation is that the corner points can be achieved using a pair of \emph{point-to-point} channel codes \cite{RimUrb96}.

Consider a rate pair \[ R_1 < \frac{1}{2} \log (1 + \frac{P}{P+N}), \ R_2 < \frac{1}{2} \log (1 + \frac{P}{N}). \]  Fix codeword length $n$ and choose a rate $R_1$ SPARC for transmitter $1$ using an $n \times M_1 L_1$ design matrix $\mathbf{A}_1$, with $M_1^{L_1} = 2^{nR_1}$ and $\mathbf{A}_1$ satisfying the specifications of Fact \ref{fact:ch_coding} for signal-to-noise ratio $\frac{P}{P+N}$. Similarly, chose a rate $R_2$ SPARC for transmitter $2$ using an $n \times M_2 L_2$ design matrix $\mathbf{A}_2$, with $M_2^{L_2} = 2^{nR_2}$ and $\mathbf{A}_2$ satisfying the specifications of Fact \ref{fact:ch_coding} for signal-to-noise ratio $\frac{P}{N}$. The codewords $X_1=\mathbf{A_1} \beta_1$ and $X_2=\mathbf{A_2} \beta_2$ chosen by the respective users, are transmitted through the channel. The non-zero values in both $\beta_1$ and $\beta_2$ are set to $\sqrt{P/L}$. The receiver obtains
\[ \mathbf{Y}=  \mathbf{A_1} \beta_1 + \mathbf{A_2} \beta_2  + \mathbf{Z} \]
and uses a successive cancellation strategy. It first decodes the message of transmitter $1$ as $\hat{\beta}_1$, effectively treating $\mathbf{A}_2 \beta_2 + \mathbf{Z}$  as noise. This step will be successful with high probability since the signal-to-noise ratio is $P/(P+N)$ and $R_1< \frac{1}{2} \log (1 + \frac{P}{P+N})$. In the second step, the receiver decodes $\beta_2$ from the residue $\mathbf{Y}-\mathbf{A} \hat{\beta}_1$ which equals $\mathbf{A_2} \beta_2  + \mathbf{Z}$ if the first step was successful, i.e., $\hat{\beta}_1 = \beta_1$. The second step will be successful with high probability since $R_2 < \frac{1}{2} \log (1 + \frac{P}{N})$. The other corner point of the rate region can be achieved by exchanging the roles of $X_1$ and $X_2$.

\subsection{The AWGN Broadcast Channel}
Consider the two-receiver scalar Gaussian broadcast channel where the  outputs of the two receivers are related to the channel input $X$ as
\[ Y_1= X + Z_1,  \quad Y_2= X+Z_2. \] $X$ has average power constraint $P$ and  the channel noises $Z_1,Z_2$ are independent zero mean Gaussian random variables with variances $N_1$ and $N_2$, respectively.  Without loss of generality, we assume that $N_2 \geq N_1$. The capacity region for this channel is given by \cite{CoverThomas}
\begin{equation}
R_1 < \frac{1}{2} \log \left(1 + \frac{\alpha P}{N_1}\right), \ \
R_2 <  \frac{1}{2} \log \left(1 + \frac{(1-\alpha)P}{\alpha P + N_2}\right). \label{eq:bc_r1r2}
\end{equation}

This capacity region is achievable through a SPARC coding scheme similar to the one for the AWGN MAC. Consider a rate pair $(R_1, R_2)$ satisfying \eqref{eq:bc_r1r2}.  Fix codeword length $n$ and choose a rate $R_1$ SPARC for transmitter $1$ using an $n \times M_1 L_1$ design matrix $\mathbf{A}_1$, with $M_1^{L_1} = 2^{nR_1}$ and $\mathbf{A}_1$ satisfying the specifications of Fact \ref{fact:ch_coding} for signal-to-noise ratio $\frac{\alpha P}{N_1}$. Similarly, chose a rate $R_2$ SPARC for transmitter $2$ using an $n \times M_2 L_2$ design matrix $\mathbf{A}_2$, with $M_2^{L_2} = 2^{nR_2}$ and $\mathbf{A}_2$ satisfying the specifications of Fact \ref{fact:ch_coding} for signal-to-noise ratio $\frac{(1-\alpha)P}{\alpha P + N_2}$. The input sequence is generated as $\bfx=\mathbf{A_1} \beta_1 + \mathbf{A_2} \beta_2$ where $\beta_1, \beta_2$ represent the messages of the two users. The non-zero values in $\beta_1$ and $\beta_2$ are set to $\sqrt{\alpha{P}/L}$ and $\sqrt{(1-\alpha){P}/L}$, respectively. The receivers obtain
\[ \mathbf{Y}_1=  \mathbf{A_1} \beta_1 + \mathbf{A_2} \beta_2  + \mathbf{Z}_1, \ \ \mathbf{Y}_2=  \mathbf{A_1} \beta_1 + \mathbf{A_2} \beta_2  + \mathbf{Z}_2.\]
Receiver $2$ decodes $\beta_2$ treating $\mathbf{A_1} \beta_1 + \mathbf{Z}_2$ as noise. This will be successful with high probability since
$R_2 <  \frac{1}{2} \log_2 \left(1 + \frac{(1-\alpha)P}{\alpha P + N_2}\right)$. Receiver $1$ can also decode $\beta_2$ with high probability since its %signal-to-noise ratio $\frac{(1-\alpha)P}{\alpha P + N_1}$ is greater than that of receiver $2$
since $N_1 \leq N_2$. Hence the residue $\mathbf{Y}_1 - \mathbf{A_2} \hat{\beta}_2$ at receiver $1$  will be equal to $\mathbf{A_1} \beta_1 + \mathbf{Z}_1$ with high probability. Receiver $1$ can then reliably decode $\beta_1$ from this residue since the rate $R_1 < \frac{1}{2} \log (1 + \frac{\alpha P}{N_1})$.

\section{Conclusion}\label{sec:conc}
The results of  \cite{AntonyML,AntonyOMP,AntonyFast,KontSPARC, RVGaussianRD12, RVGaussianFeasible} showed that the sparse regression framework can be used to design rate-optimal codes for point-to-point source and channel coding with computationally efficient encoders and decoders. In this paper, we showed how these source and channel codes can be combined to implement random binning and superposition. These two techniques have been  fundamental ingredients of rate-optimal coding schemes for a wide range of problems in network information theory. The next goal is a precise performance analysis of the computationally feasible versions of the coding schemes presented here. Constructing a library of efficient sparse regression modules to perform source coding, channel coding, binning and superposition will pave the way for fast, rate-optimal codes for several network problems such as  multiple description coding, lossy distributed source coding, interference channels and relay channels.

%\textbf{random binning, superposition are the workhorses of multi-user information theory - how we can realize them with SPARCs}

\IEEEtriggeratref{31}

\bibliographystyle{ieeetr}
\bibliography{proposal}
\end{document}